\documentclass[preprint,noshowpacs,preprintnumbers,amsmath,amssymb,superscriptaddress,nofootinbib]{revtex4}
\usepackage{graphicx,color}
\usepackage[]{hyperref}
\usepackage{amsmath,amssymb}
\usepackage{url}
\usepackage{bbold}

\newcommand{\hs}{\hspace*{0.3cm}}

\newcommand{\eq}[1]{Eq.~(\ref{#1})}
\newcommand{\bib}[1]{Ref.~\cite{#1}}
\newcommand{\refs}[1]{Refs.~\cite{#1}}

\newcommand{\tab}[1]{Table~\ref{#1}}
\newcommand{\sect}[1]{Section~\ref{#1}}

\newcommand{\appen}[1]{Appendix~\ref{#1}}

\newcommand{\be}{\begin{equation}}
\newcommand{\ee}{\end{equation}}
\newcommand{\bea}{\begin{eqnarray}}
\newcommand{\eea}{\end{eqnarray}}

\newcommand{\crn}{\nonumber \\}

\newcommand{\fr}{\frac}
\newcommand{\bc}{\begin{center}}
\newcommand{\ec}{\end{center}}

\newcommand {\ba}{\begin{array}}
\newcommand {\ea}{\end{array}}
\newcommand{\ben}{\begin{enumerate}}
\newcommand{\een}{\end{enumerate}}

\begin{document}

\preprint{IFIRSE-TH-2018-5}

\title{On the triplet anti-triplet symmetry in 3-3-1 models}
\author{Le Tho Hue}\email{lthue@iop.vast.ac.vn}
\affiliation{\small
Institute for Research and Development, Duy Tan University, 550000 Da Nang, Vietnam
}
\affiliation{Institute of Physics, Vietnam Academy of Science and Technology, 10 Dao Tan, Ba
Dinh, Hanoi, Vietnam }
\author{Le Duc Ninh}\email{ldninh@ifirse.icise.vn}
\affiliation{Institute For Interdisciplinary Research in Science and Education, 
ICISE, 590000 Quy Nhon, Vietnam}

\begin{abstract}
We present a detailed discussion of the triplet anti-triplet symmetry in 3-3-1 models. 
The full set of conditions to realize this symmetry is provided, which includes in particular 
the requirement that the two vacuum expectation values of the two scalar triplets responsible for making the $W$ and $Z$ bosons massive 
must be interchanged. We apply this new understanding 
to the calculation of processes that have a $Z-Z'$ mixing. 
\end{abstract}
\pacs{{\bf Last updated: \today}
%%12.60.Fr  Extensions of electroweak Higgs sector,
}
\maketitle
%%%%%%%%%%%%%%%%%%%

\section{Introduction}
\label{sec:intro}
Interesting extensions of the Standard Model (SM), based on the local gauge group $SU(3)_C\otimes SU(3)_L\otimes U(1)_X$ (3-3-1), have been widely studied (see \bib{Singer:1980sw} 
and references therein, see also Refs.~\cite{Valle:1983dk,Pisano:1991ee,Foot:1992rh,Frampton:1992wt,Foot:1994ym} for similar models 
but with lepton-number violation). 
Fermions are typically organized into triplets and anti-triplets of $SU(3)_L$ in three generations. 
We therefore have two possible choices, either to put leptons in triplets or anti-triplets. 

The 3-3-1 models can be classified using the parameter $\beta$ defined via the electric charge operator
\bea
Q = T_3 + \beta T_8 + X \mathbb{1},
\label{eq:charge_Q}
\eea
where $T_3$, $T_8$ are the diagonal $SU(3)$ generators and $X$ is a new quantum charge of the group $U(1)_X$. 

For a given $\beta \neq 0$, the model with the leptons in triplets is different from the one where 
the leptons are in anti-triplets, because the two models have different electric-charge spectra. Also, 
for a given assignment of fermionic representation, changing the sign of $\beta$ will lead to a 
different electric-charge spectrum. What happens if we switch both simultaneously? 
We call this 
\bea
\beta \to -\beta \quad \text{and} \quad \text{triplets} \leftrightarrow \text{anti-triplets},
\label{t_at_trans}
\eea
triplet anti-triplet transformation for short. This transformation applies also to quarks, i.e. quark triplets $\to$ 
quark anti-triplets and quark anti-triplets $\to$ 
quark triplets.   
In the special case of $\beta = 0$ \cite{Hue:2015mna}, this becomes truly a triplet anti-triplet transformation literally. 

Using \eq{eq:charge_Q} one can easily see that the electric-charge spectrum is invariant. Therefore, at first glance, 
we expect that physics must be the same. This was noted e.g. in \bib{Descotes-Genon:2017ptp} (see the sentence after Eq. (2.4) therein) and at the end of 
Section II in \bib{Richard:2013xfa}. We think that this is well recognized in the 3-3-1 model community. 

However, this understanding is put into question, at least for us, after reading the paper of Buras, De Fazio and Girrbach-Noe \cite{Buras:2014yna} (see also \cite{Buras:2015kwd}) 
where there are indications that this symmetry is broken by $Z-Z'$ mixing, which depends on the sign of $\beta$ but, apparently, not 
on the fermionic representation at tree level. More specifically, the $Z-Z'$ mixing angle is given, in $M_Z \ll M_{Z'}$ approximation, by \cite{Buras:2014yna}
\bea
\sin\xi = \fr{c_W^2}{3\sqrt{1-(1+\beta^2)s^2_W}}\left(3\beta\fr{s^2_W}{c^2_W}+\sqrt{3}a\right)\fr{M_Z^2}{M_{Z'}^2},
\label{Sin_ZZ'_Buras}
\eea 
where $s^2_W = \sin^2\theta_{W}$, $c^2_W = 1 - s^2_W$ with $\theta_W$ being the weak-mixing angle and 
\bea
a=\fr{v^2_1 - v^2_2}{v^2_1+v^2_2},
\label{a_Buras}
\eea
where $v_1$ and $v_2$ (called $v_\eta$ and $v_\rho$ in \bib{Buras:2014yna}, respectively) are the vacuum expectation values of the two Higgs triplets responsible for making $W^\pm$ and $Z$ bosons massive. 
From \eq{Sin_ZZ'_Buras}, one can see that the absolute value of the mixing angle changes under $\beta \to -\beta$. 
The authors of \bib{Buras:2014yna} further pointed out that $\sin\xi$ does not depend on whether the leptons are assigned in triplets 
or anti-triplets. This seems to indicate that the $Z-Z'$ mixing breaks the triplet anti-triplet symmetry, see the discussion around Eq. (2.16) of \bib{Buras:2014yna}. Many figures in \bib{Buras:2014yna} also seem to support this conclusion.

We notice, however, one missing ingredient in the above discussion. Namely, the parameter $a$ defined in \eq{a_Buras} changes sign under the 
triplet anti-triplet transformation. The authors of \bib{Buras:2014yna} did not see this because they chose $a$ to be an input parameter 
and kept it unchanged under the transformation. If we instead choose the charged gauge-boson masses 
as independent input parameters and calculate $a$ from them, then we will see that the value of $a$ changes sign. 
{\em This is the main point of this letter, which, to the best of our knowledge, has not been noted in the literature.} 
The choice of the charged-gauge boson masses as input parameters 
is natural as this is directly related to physical observables. 
\bib{Buras:2014yna} focused on the neutral gauge bosons and did not touch the charged gauge-boson masses, hence this important point was 
missed out. With this new piece of information, we will see that $\sin\xi$ can only change sign under the triplet anti-triplet transformation. 

In trying to solve this puzzle, we have realized that there exists no detailed discussion of the 
triplet anti-triplet symmetry in the literature apart from some brief remarks as above noted. 
Since this is an important issue in 3-3-1 models, we think it can be useful to 
show in detail how this symmetry works. 
We have found that the actual implementation of this symmetry in practice requires not only 
a careful attention to the input parameter scheme as above noted but also possible sign flips in many places 
in the Feynman rules and in book-keeping parameters. We will also show that the full definition of the triplet anti-triplet transformation 
is more complicated than \eq{t_at_trans} and changing the sign of the parameter $a$. This is easy to see because the full Lagrangian depends 
also on many other parameters, which may also flip signs or interchange under the transformation. 

There is another issue related to the comparison with \bib{Buras:2014yna}. Indeed, \bib{Buras:2014yna} provided 
results for two models called $F_1(\beta,a)$ and $F_2(-\beta,a)$, related by the transformation \eq{t_at_trans}. For each model, 
results for different values $a = 0,\pm 12/13$ are also given. Numerical results of \bib{Buras:2014yna}, see Fig. 4 and Fig. 5 therein, 
show that $F_1(\beta,a)$ and $F_2(-\beta,-a)$ are not the same. This is very surprising to us because we expect them to be identical 
according to the triplet anti-triplet symmetry. We have discussed this issue with the authors of \bib{Buras:2014yna}, but, unfortunately, 
no conclusive finding has been reached. Our investigation has led us to the 
conclusion that there seems to be an issue with the sign of the couplings between the $Z'$ and 
the leptons in the model $F_2$. 

The paper is organized as follows. In \sect{sect:two_models}, we discuss the two models related by the transformation 
\eq{t_at_trans} and provide the full set of conditions for them to be identical. In \sect{sect:apply}, we make application 
to the processes with a $Z-Z'$ mixing and perform some crosschecks 
with \bib{Buras:2014yna} and other papers. Conclusions are given in \sect{conclusions}. 
In \appen{appen:Feynman_rules} we provide details on the calculation of the $Z-Z'$ mixing and of the 
couplings between the $Z$, $Z'$ gauge bosons to the leptons in the two models with a general sign convention for the 
$Z'$ field definition. 

%%%
\section{Two identical models} 
\label{sect:two_models}
In this section we consider two 3-3-1 models denoted $M_1$ and $M_2$, related by 
the triplet anti-triplet transformation defined by \eq{t_at_trans}. We note that 
\eq{t_at_trans} is not enough to make the two models identical, because 
the physical results depend also on the values of other input parameters such as masses, 
mixing and coupling parameters. Since we impose here that the two 
models are identical, there must be relations between the parameters of the two models. 
These relations can be found by comparing the two Lagrangians.   

The parameter $\beta$ will be denoted $\beta_1$ and $\beta_2$ for the two models, respectively. 
We will use the indices $m,n = 1,2$ to distinguish the models.
 
The model $M_1$ is 
defined as follows. 
Left-handed leptons are assigned into anti-triplets and right-handed leptons are 
singlets: 
\bea && L_{aL}=\left(
       \begin{array}{c}
         e_a \\
         -\nu_{a} \\
         E_a \\
       \end{array}
     \right)_L \sim \left(3^*, -\frac{1}{2}+\frac{\beta_1}{2\sqrt{3}}\right), \hs a=1,2,3,\crn
     && e_{aR}\sim   \left(1, -1\right)  , \hs \nu_{aR}\sim \left(1, 0\right) ,\hs E_{aR} 
\sim   \left(1, -\frac{1}{2}+\frac{\sqrt{3}\beta_1}{2}\right).  \label{lep_F1}\eea
The model includes three right-handed neutrinos $\nu_{aR}$. 
The leptons $E^a_{L,R}$ can be new particles or charge-conjugated states of the SM leptons.  
In the following, we will assume, without loss of generality, $E_a$ to be new leptons.  
The numbers in the parentheses are to label the representations of $SU(3)_L$ and $U(1)_X$ groups. 
Note that we have $Q=X$ for singlets. 

Anomaly cancellation requires that the number of triplets and anti-triplets must be equal. 
Since quarks come in three colors, this means that one family of quarks must be in anti-triplet 
and the other two families are in triplets or vice versa. 
This implies two choices, the leptons are either put in triplets 
or in anti-triplets. Because Feynman rules for the quarks are similar to those 
for the leptons, we will ignore the quarks and focus on the leptons in the following. 

For $M_2$, the left-handed leptons are put in triplets as 
\bea && L_{aL}=\left(
       \begin{array}{c}
         \nu_{a} \\
         e_a \\
         E_a \\
       \end{array}
     \right)_L \sim \left(3~, -\frac{1}{2}-\frac{\beta_2}{2\sqrt{3}}\right), \hs a=1,2,3,\crn
     && e_{aR}\sim   \left(1~, -1\right)  , \hs \nu_{aR}\sim  \left(~1~, 0\right) ,\hs E_{aR} 
\sim   \left(~1~, -\frac{1}{2}-\frac{\sqrt{3}\beta_2}{2}\right).  \label{lep_F2}\eea
Note that the positions of $\nu_{aL}$ and $e_{aL}$ have been interchanged to make the Feynman rules 
for the SM particles the same as those in the SM. 
Requiring that the electric charges of $E_a$ in both models are the same leads to
\bea
\beta_1 = -\beta_2. \label{eq:relation_beta}
\eea 

We use the same convention for the Lagrangian of both models as
\begin{align}
\mathcal{L}_\text{lepton} &= \bar{L}_{aL} i\gamma^\mu D_\mu L_{aL} + \bar{e}_{aR} i\gamma^\mu D_\mu e_{aR} 
+ \bar{\nu}_{aR} i\gamma^\mu D_\mu \nu_{aR},\label{eq:L_lep}\\
\mathcal{L}^\text{kinetic}_\text{scalar}&= (D_{\mu} \Phi_i)^\dagger (D^\mu \Phi_i), \label{eq:L_kin}\\
\mathcal{L}_{1,\text{Yukawa}}&= -Y^e_{1,ab} \overline{L}_{aL} \Phi_1^*e_{bR}- Y^\nu_{1,ab} \overline{L}_{aL} \Phi_2^*\nu_{bR} - Y^E_{1,ab} \overline{L}_{aL} \Phi_3^*E_{bR}+\mathrm{h.c.},\label{eq:L_Yuk_1}\\
\mathcal{L}_{2,\text{Yukawa}}&= -Y^e_{2,ab} \overline{L}_{aL} \Phi_2^*e_{bR}- Y^\nu_{2,ab} \overline{L}_{aL} \Phi_1^*\nu_{bR} - Y^E_{2,ab} \overline{L}_{aL} \Phi_3^*E_{bR}+\mathrm{h.c.},
\label{eq:L_Yuk_2}
\end{align}
where $i=1,2,3$ to denote the three scalar multiplets. The Yukawa Lagrangians are written for both models explicitly.  
The Yukawa couplings are the same for both models, namely 
\bea
Y^l_{1,ab} = Y^l_{2,ab},\quad l = e,\nu, E. 
\label{eq:relation_Yukawa}
\eea  
The covariant derivative reads
\begin{align}
D_\mu^\text{triplet} &\equiv \partial_{\mu}-i g T_s W^{s}_{\mu}-i g_X X T_9 X_{\mu},\crn  
D_\mu^\text{anti-triplet} &\equiv \partial_{\mu}+i g (T_s)^T W^{s}_{\mu}-i g_X X T_9X_{\mu},\crn
D_\mu^\text{singlet} &\equiv \partial_{\mu}-i g_X X T_9 X_{\mu},   
\label{coderivative}
\end{align}
where $T_s =\lambda_s/2$ with $s=1,\ldots,8$ and $\lambda_s$ being Gell-Mann matrices, $T_9=1/\sqrt{6}$, $g$ and $g_X$ are coupling constants corresponding to the two groups $SU(3)_L$ and $U(1)_X$, respectively. 
Their values are the same in both models. 
We further define
\begin{align} \mathcal{W}^\text{triplet}_\mu &\equiv W^s_{\mu}T_s=\frac{1}{2}\left(
                     \begin{array}{ccc}
                       W^3_{\mu}+\frac{1}{\sqrt{3}} W^8_{\mu}& \sqrt{2}W^+_{\mu} &  \sqrt{2}Y^{+A_m}_{\mu} \\
                        \sqrt{2}W^-_{\mu} &  -W^3_{\mu}+\frac{1}{\sqrt{3}} W^8_{\mu} & \sqrt{2}V^{+B_m}_{\mu} \\
                       \sqrt{2}Y^{-A_m}_{\mu}& \sqrt{2}V^{-B_m}_{\mu} &-\frac{2}{\sqrt{3}} W^8_{\mu}\\
                     \end{array}
                   \right),\crn
\mathcal{W}^\text{anti-triplet}_\mu &\equiv -W^s_{\mu}(T_s)^T = - (\mathcal{W}^\text{triplet}_\mu)^T, 
  \label{wata}\end{align}
where $m=1,2$ and we have defined the mass eigenstates of the charged gauge bosons as
\bea W^{\pm}_{\mu}=\frac{1}{\sqrt{2}}\left( W^1_{\mu}\mp i W^2_{\mu}\right),\crn
Y^{\pm A_m}_{\mu}=\frac{1}{\sqrt{2}}\left( W^4_{\mu}\mp i W^5_{\mu}\right),\crn
V^{\pm B_m}_{\mu}=\frac{1}{\sqrt{2}}\left( W^6_{\mu}\mp i W^7_{\mu}\right).
   \label{gbos}\eea 
The electric charges of the gauge bosons are calculated as 
\bea
A_m=\fr{1}{2}+\beta_m\fr{\sqrt{3}}{2}, \quad 
B_m=-\fr{1}{2}+\beta_m\fr{\sqrt{3}}{2}\label{charge_gauge}.\eea
We see clearly that $\beta_2 = -\beta_1$ is 
equivalent to $B_m = -A_n$ with $m\neq n$. 

\eq{eq:L_Yuk_1} and \eq{eq:L_Yuk_2}
require that $\Phi_i$ are triplets in $M_1$ and anti-triplets in $M_2$. 
This is just a matter of convention and we can e.g. change $\Phi_i$ to be 
triplets in $M_2$ by removing the complex conjugation in \eq{eq:L_Yuk_2}. 
For $M_1$ we have
  \bea && \Phi_3=\left(
              \begin{array}{c}
                \phi_3^{+A_1} \\
                \phi_3^{+B_1} \\
                \phi_3^0 \\
              \end{array}
            \right)\sim \left(3~, \frac{\beta_1}{\sqrt{3}}\right), \hs  \Phi_2=\left(
              \begin{array}{c}
                \phi_2^+ \\
                \phi_2^0 \\
                \phi_2^{-B_1} \\
              \end{array}
            \right)\sim \left(3~, \frac{1}{2}-\frac{\beta_1}{2\sqrt{3}}\right)\crn
  && \Phi_1=\left(
              \begin{array}{c}
                \phi_1^0 \\
                \phi_1^- \\
                \phi_1^{-A_1} \\
              \end{array}
            \right)\sim \left(3~, -\frac{1}{2}-\frac{\beta_1}{2\sqrt{3}}\right).
    \label{scalar_F1}
  \eea
And for $M_2$
  \bea && \Phi_3=\left(
              \begin{array}{c}
                \phi_3^{-A_2} \\
                \phi_3^{-B_2} \\
                \phi_3^0 \\
              \end{array}
            \right)\sim \left(3^*~, \frac{-\beta_2}{\sqrt{3}}\right), \hs  \Phi_2=\left(
              \begin{array}{c}
                \phi_2^- \\
                \phi_2^0 \\
                \phi_2^{+B_2} \\
              \end{array}
            \right)\sim \left(3^*~, -\frac{1}{2}+\frac{\beta_2}{2\sqrt{3}}\right)\crn
  && \Phi_1=\left(
              \begin{array}{c}
                \phi_1^0 \\
                \phi_1^+ \\
                \phi_1^{+A_2} \\
              \end{array}
            \right)\sim \left(3^*~, \frac{1}{2}+\frac{\beta_2}{2\sqrt{3}}\right).
    \label{scalar_F2}
  \eea
The scalar fields develop vacuum expectation values (VEV) defined as 
  \bea&& \langle  \Phi_3\rangle=\frac{1}{\sqrt{2}}\left(
              \begin{array}{c}
                0 \\
                0 \\
                v_{m,3} \\
              \end{array}
            \right), \hs \langle  \Phi_2 \rangle =\frac{1}{\sqrt{2}}\left(
              \begin{array}{c}
                0 \\
                v_{m,2} \\
                0 \\
              \end{array}
            \right),\hs \langle   \Phi_1 \rangle= \frac{1}{\sqrt{2}}\left(
              \begin{array}{c}
                v_{m,1} \\
                0 \\
                0 \\
              \end{array}
            \right). \label{vevhigg_F1}\eea

We now discuss gauge boson masses. From \eq{eq:L_kin} we get 
for charged gauge bosons
\bea
m_W^2 = \fr{g^2}{4}(v_{m,1}^2 + v_{m,2}^2),\quad 
m^2_{Y^{\pm A_m}} = \frac{g^2}{4}(v_{m,3}^2+v_{m,1}^2),\quad 
m^2_{V^{\pm B_m}}=\frac{g^2}{4}(v_{m,3}^2+v_{m,2}^2).
\label{eq:mass_gauge_charge}
\eea
As noticed, under the transformation $\beta_2 = -\beta_1$ we have $A_m = -B_n$ with $m\neq n$, 
hence these equations lead to 
\bea
v_{1,3} = v_{2,3}, \quad v_{m,1} = v_{n,2},
\label{eq:relation_vevs}
\eea 
which come from the condition that {\em both models must have the same charged gauge bosons 
(i.e. same electric charges and same masses).} It is straight forward to see that the neutral gauge bosons 
have the same masses in both models, see \appen{appen:Feynman_rules}. 
It is convenient to introduce the following parameter
\bea
a_m=\fr{v^2_{m,1}-v^2_{m,2}}{v^2_{m,1}+v^2_{m,2}}, 
\label{eq:def_a}
\eea
which was defined in \bib{Buras:2014yna} and was mentioned in the introduction. 
Because of \eq{eq:relation_vevs}, we have 
\bea
a_1 = -a_2.
\label{eq:relation_a}
\eea

We now consider the scalar potentials, which read 
\bea V_{m}&=&\mu_{m,1}^2 \Phi_1^{\dagger}\Phi_1+\mu_{m,2}^2\Phi_2^{\dagger}\Phi_2+\mu_{m,3}^2\Phi_3^{\dagger}\Phi_3
+\lambda_{m,1} \left(\Phi_1^{\dagger}\Phi_1\right)^2
+\lambda_{m,2}\left(\Phi_2^{\dagger}\Phi_2\right)^2
+\lambda_{m,3}\left(\Phi_3^{\dagger}\Phi_3\right)^2\crn
&+& \lambda_{m,12}(\Phi_1^{\dagger}\Phi_1)(\Phi_2^{\dagger}\Phi_2)
+\lambda_{m,13}(\Phi_1^{\dagger}\Phi_1)(\Phi_3^{\dagger}\Phi_3)
+\lambda_{m,23}(\Phi_2^{\dagger}\Phi_2)(\Phi_3^{\dagger}\Phi_3)\crn
&+&\tilde{\lambda}_{m,12} (\Phi_1^{\dagger}\Phi_2)(\Phi_2^{\dagger}\Phi_1) 
+\tilde{\lambda}_{m,13} (\Phi_1^{\dagger}\Phi_3)(\Phi_3^{\dagger}\Phi_1)
+\tilde{\lambda}_{m,23} (\Phi_2^{\dagger}\Phi_3)(\Phi_3^{\dagger}\Phi_2)\crn
&+&\sqrt{2} f_{m}\left(\epsilon_{ijk}\Phi_1^i\Phi_2^j\Phi_3^k +\mathrm{h.c.} \right), \quad m=1,2.
\label{eq:scalar_potential}\eea
In order to have the relations in \eq{eq:relation_vevs} we must have, with $m\neq n$,
\begin{align}
&\mu^2_{m,1} = \mu^2_{n,2}, \quad \mu^2_{1,3} = \mu^2_{2,3},
\quad \lambda_{m,1} = \lambda_{n,2}, \quad \lambda_{1,3} = \lambda_{2,3},\crn 
&\lambda_{1,12} = \lambda_{2,12}, \quad \lambda_{m,13} = \lambda_{n,23}, 
\quad \tilde{\lambda}_{1,12} = \tilde{\lambda}_{2,12}, \quad \tilde{\lambda}_{m,13} = \tilde{\lambda}_{n,23}, 
\quad f_{1} = f_{2}.
\label{eq:relation_potential}  
\end{align}
With these relations it is straight forward to see that the Higgs mass spectra of the two models are identical and 
the vertices of pure scalar, scalar-fermion, scalar gauge boson interactions are the same. 

In summary, the two models $M_1$, where the leptons are organized in anti-triplets, and $M_2$, where 
the leptons are in triplets, are equivalent if, besides identical gauge couplings, 
the relations \eq{eq:relation_beta}, \eq{eq:relation_Yukawa}, and \eq{eq:relation_potential} are satisfied. 
The important relation \eq{eq:relation_a} is a consequence of \eq{eq:relation_potential}. 
We therefore remark that the conditions in 
\eq{t_at_trans} are necessary but not sufficient to realize the 
triplet anti-triplet symmetry. 

\section{Application to neutral-current processes}
\label{sect:apply}
For the following discussion, 
it is useful to define the models as follows
\bea
M_1 = M(3^*,\beta_1,a_1), \quad M_2 = M(3,\beta_2,a_2),
\eea
where the first argument specifies the representation for the leptons. 
Of course, those three arguments are not enough to define a model, but they will be 
enough for our purpose in this section, assuming that the gauge and Yukawa couplings are the same, 
and the parameter $a$ represents the parameters of the scalar potential. 
With these assumptions, we have 
\bea
M(3^*,\beta,a) = M(3,-\beta,-a),
\eea 
as a simplified way of expressing the triplet anti-triplet symmetry. 

In \bib{Buras:2014yna}, two models are discussed $F_1(a) = M(3^*,\beta,a)$ and $F_2(a) = M(3,-\beta,a)$. 
\bib{Buras:2014yna} introduced also the parameter $\tan\bar{\beta} = v_{m,2}/v_{m,1}$, which 
can be related to $a$ via
\bea
a = \fr{1-\tan^2\bar{\beta}}{1+\tan^2\bar{\beta}}.
\eea
The transformation $a \to -a$ is therefore equivalent to $\tan\bar{\beta} \to 1/\tan\bar{\beta}$.
 
From the above discussion, we can now see clearly that $F_1(a)$ and $F_2(a)$ are not equivalent, 
leading (unsurprisingly) to the fact that the results for $F_1(a)$ and for $F_2(a)$ are not the same if $a\neq 0$.
 
However, the results of model $F_2(-a)$ are also provided in \bib{Buras:2014yna} and they are not the same 
as those of $F_1(a)$. This is unexpected because the triplet anti-triplet symmetry suggests that they 
should be the identical. 
The calculation of \bib{Buras:2014yna} involves two neutral currents mediated by $Z_1$ and $Z_2$ particles. 
These mass eigenstates are related to the $Z$ and $Z'$ states as
\begin{align}
Z_1^\mu = \cos\xi Z^\mu + \sin\xi Z'^\mu, \quad Z_2^\mu = -\sin\xi Z^\mu + \cos\xi Z'^\mu .
\end{align}
More details are provided in \appen{appen:Feynman_rules}. The amplitude squared therefore depends 
on the sign of the $ffZ$ and $ffZ'$ couplings and also on the sign of $\tan\xi$, because of the $Z-Z'$ 
interference terms. Since the convention of $\cos\xi > 0$ is usually chosen, we thus have to pay attention 
to the sign of the couplings and of $\sin\xi$. 

We have made an investigation into \bib{Buras:2014yna} and come to the conclusion that there seems 
to be a sign issue in the $llZ'$ couplings of $F_2$. We have performed the following checks. 
\begin{itemize}
\item For model $F_1$, we agree with Table 1 of \bib{Buras:2014yna}. 
\item For model $F_2$, we agree with \bib{Martinez:2014lta}. 
\item For model $F_2$, we agree with \bib{Buras:2014yna} on $\sin\xi$ and can 
reproduce the Table 2 of \bib{Buras:2014yna} {\em if} a minus sign is added to the $llZ'$ couplings 
\footnote{Additionally, in \bib{Buras:2014yna},  
there seems to be sign typos at the $R_V^{\mu\mu}$ result for $\beta = 2/\sqrt{3}$ and $\tan\bar{\beta}=5$ 
in Table 1, and at the $W_A^{\mu\mu}$ result for $\beta = -2/\sqrt{3}$ and $\tan\bar{\beta}=5$ in Table 2.}. 
However, if the correct sign is used, the results change because of the $Z-Z'$ interference terms. 
\end{itemize}

Note that the sign of $\sin\xi$ in \bib{Martinez:2014lta} agrees with \bib{Buras:2014yna} and 
also with this paper. The $llZ$, $llZ'$ couplings of \bib{Martinez:2014lta} are the same as in 
\bib{CarcamoHernandez:2005ka} and agree with this paper. \bib{Buras:2014yna} and 
\bib{Martinez:2014lta} did not mention whether they agree on the $llZ$, $llZ'$ couplings.  

To facilitate comparisons, our results for the $llZ$, $llZ'$ couplings 
and for $\sin\xi$ are provided in \appen{appen:Feynman_rules}. All these findings have been communicated 
to the authors of \bib{Buras:2014yna}. 

\section{Conclusions}
\label{conclusions}
In this work we have pointed out that the recognized triplet anti-triplet symmetry in 3-3-1 models should include 
a sign change in the parameter $a = (v_1^2-v_2^2)/(v_1^2 + v_2^2)$, besides the well-known sign change in the parameter $\beta$ and 
changing from triplets to anti-triplets and vice versa. 
We have shown that the full transformation is more complicated than that and attention has to be paid 
to the input parameter scheme and also to the parameters of the scalar potential. The transformations of those parameters have been provided. 

We have applied the new understanding to the processes with a $Z-Z'$ mixing and in particular to the 
calculations of \bib{Buras:2014yna}. We have found a possible sign issue with 
the couplings between the $Z'$ and the leptons in the model where the leptons are put 
in the triplet representation. 

\section*{Acknowledgments}
We would like to thank Andrzej Buras and Fulvia De Fazio for discussions. 
We are grateful to Julien Baglio for his careful reading of the manuscript 
and for his helpful comments.   
This research is funded by the Vietnam
National Foundation for Science and Technology Development (NAFOSTED)
under grant number 103.01-2017.78. LDN acknowledges the support from
DAAD to perform the final part of this work at the University of
T\"ubingen under a scholarship. He thanks the members of the 
Institut f\"{u}r Theoretische Physik for their hospitality. 

\appendix

\section{$Z-Z'$ mixing}
\label{appen:Feynman_rules}
We consider here the neutral gauge bosons. For both models defined in \sect{sect:two_models}, 
we will introduce intermediate fields 
of $Z_\mu$ and $Z'_\mu$ and the final physical fields will be $A^\mu$ (the photon), $Z_{1}^\mu$ 
and $Z_{2}^\mu$. We will see that the masses of these physical states are 
the same in both models. 

The symmetry breaking pattern is 
\bea
SU(3)_L\otimes U(1)_X\xrightarrow{v_{m,3}} SU(2)_L\otimes U(1)_Y\xrightarrow{v_{m,1},v_{m,2}} U(1)_Q.
\eea
The basis of the neutral gauge bosons transforms correspondingly as
\bea 
X_{\mu},\, W^8_{\mu},\, W^3_{\mu} \xrightarrow{\theta_{m,331}} Z'_{\mu},\, B'_{\mu},\, W^3_{\mu} 
\xrightarrow{\theta_W} Z'_{\mu},\, Z_{\mu},\, A_{\mu}. 
\eea
We have
\begin{align}
A_\mu    &= s_W W^3_\mu + c_W(s_{m,331}X_\mu + c_{m,331}W^8_\mu),\label{eq:Amu}\\
Z_\mu    &= c_W W^3_\mu - s_W(s_{m,331}X_\mu + c_{m,331}W^8_\mu),\label{eq:Zmu}\\
Z'_{\mu} &= h_m(-s_{m,331} W^8_\mu + c_{m,331}X_\mu),\label{eq:Zpmu}
\end{align}
with $h_m = \pm 1$ being a sign convention for $Z'_{\mu}$, and
\begin{align}
s_{m,331} &= \sin\theta_{m,331} = \fr{g}{\sqrt{X_g}}, \quad c_{m,331} = \cos\theta_{m,331} = \fr{g_X\beta_m}{\sqrt{6X_g}}, 
\quad X_g = g^2 + \fr{g_X^2\beta_m^2}{6},\label{eq:angle_331}\\
s_W &= \sin\theta_W = \fr{g_1}{\sqrt{g^2 + g_1^2}}, \quad c_W = \cos\theta_W = \fr{g}{\sqrt{g^2 + g_1^2}}, 
\quad g_1 = \fr{gg_X}{\sqrt{6X_g}},\label{eq:angle_W}\\
g_X^2 &= \fr{6g^2 s^2_W}{1-(1+\beta_m^2)s^2_W}.\label{eq:relation_gX_g}
\end{align}
If we choose $h_1 = +1$ for model $M_1$, as in \bib{Carlucci:2013owm}, then 
we agree with \bib{Buras:2014yna}. For $M_2$, we agree with \bib{Martinez:2014lta} 
where the convention $h_2 = +1$ is used. 

The mass matrix in the basis of $(A_\mu, Z_\mu, Z'_\mu)$ for both models is
\bea
M^2_{AZZ'}=
\left(\begin{array}{ccc}
                0 & 0 & 0 \\
                0 & m^2_{22} & m^2_{23} \\
                0 & m^2_{23} & m^2_{33} \\
              \end{array}
\right),\label{eq:matrix_AZZp}
\eea
with
\bea
m^2_{22} &=& \fr{m^2_W}{c^2_W}, \quad m^2_{23}=-\fr{h_m g_X m^2_W}{3\sqrt{2}gs_W}(a_m+\sqrt{3}\beta_m t^2_W),\crn
m^2_{33} &=& \fr{X_g v_{m,3}^2}{3} + \fr{g_X^2 m^2_W}{18g^2t^2_W}(1+3 t^4_W \beta_m^2 + 2\sqrt{3}t^2_W\beta_ma_m),
\eea
where
\bea
a_m=\fr{v^2_{m,1}-v^2_{m,2}}{v^2_{m,1}+v^2_{m,2}}.
\eea
Because of \eq{eq:relation_vevs}, we have 
\bea
a_1 = -a_2,
\eea
confirming what we wrote in the introduction. 

The two eigenvalues read
\bea
m^2_{Z_1} = \fr{m^2_{22}+m^2_{33}-\sqrt{\Delta}}{2}, \quad
m^2_{Z_2} = \fr{m^2_{22}+m^2_{33}+\sqrt{\Delta}}{2}, \quad
\Delta = (m^2_{22}-m^2_{33})^2 + 4m^4_{23}. 
\eea
We now introduce the $Z-Z'$ mixing discussed in the introduction
\bea
\left(\begin{array}{c}
                Z_1^\mu \\
                Z_2^\mu \\
              \end{array}
\right)
=
\left(\begin{array}{cc}
                c_{m,\xi} & s_{m,\xi} \\
                -s_{m,\xi} & c_{m,\xi} \\
              \end{array}
\right)
\left(\begin{array}{c}
                Z^\mu \\
                Z'^{\mu} \\
              \end{array}
\right),
\label{eq:rotation_ZZ'}
\eea
with 
\bea 
s_{m,\xi} = -\fr{m^2_{23}}{\sqrt{(m^2_{Z_2}-m^2_{22})^2 + m^4_{23}}}, 
\quad c_{m,\xi} = \fr{m^2_{Z_2}-m^2_{22}}{\sqrt{(m^2_{Z_2}-m^2_{22})^2 + m^4_{23}}}.
\label{eq:cos_sin_xi}
\eea
We note here that the sign of $\tan\xi_m = s_{m,\xi}/c_{m,\xi}$ is completely determined from 
the matrix in \eq{eq:matrix_AZZp}, but not the sign of $s_{m,\xi}$ or $c_{m,\xi}$. 
Here we choose the $c_{m,\xi} > 0$ convention, so that the $Z_1$ couplings become 
identical to the $Z$ couplings in the limit $\xi \to 0$. In this convention, 
the sign of $s_{m,\xi}$ is determined as in \eq{eq:cos_sin_xi}. 

Comparing $M_1$ against $M_2$ we see that $m^2_{22}$ and $m^2_{33}$ are identical but $m^2_{23}$ 
can be different in sign depending on the convention of $h_m$. 
The physical masses $m^2_{Z_i}$ are therefore the same in both models. 
For the mixing angles we have
\begin{align}
s_{1,331} = s_{2,331}, \quad c_{1,331} = -c_{2,331}, \quad c_{1,\xi} &= c_{2,\xi}, 
\label{eq:compare_331_xi} 
\end{align}
while $s_{1,\xi} = \pm s_{2,\xi}$ depending on the sign convention of $h_m$. 
$c_W$ and $s_W$ are chosen to be the same in both models. 

In the case of $v_{m,3}\gg v_{m,1}, v_{m,2}$ we have $m_{33}\gg m_{22}, m_{23}$ and
\bea
s_{m,\xi} \approx -\fr{m^2_{23}}{m^2_{33}} = \fr{h_m c^2_W}{3\sqrt{1-(1+\beta^2_m)s^2_W}}(3\beta_m t^2_W + \sqrt{3}a_m)\fr{m_{22}^2}{m_{33}^2}.
\label{sin_xi_approx}
\eea
This result agrees with \refs{Buras:2014yna,Martinez:2014lta} if we choose $h_1 = h_2 = +1$. 

To calculate neutral currents couplings to the leptons we need the diagonal entries of the covariant 
derivative. Writing $D^\text{diag}_\mu = \partial_\mu + \hat{D}_\mu$, in the original basis $(X_\mu,W^8_\mu,W^3_\mu)$ 
we have
\begin{align}
\hat{D}^\text{lepton,L}_{m,\mu} = -i\text{diag} &\left(\fr{gs_m}{2}W^3_\mu + \fr{gs_m}{2\sqrt{3}}W^8_\mu+\fr{g_X}{\sqrt{6}}(-\fr{1}{2}-\fr{\beta_ms_m}{2\sqrt{3}})X_\mu \right. , \crn
&\left. -\fr{gs_m}{2}W^3_\mu + \fr{gs_m}{2\sqrt{3}}W^8_\mu+\fr{g_X}{\sqrt{6}}(-\fr{1}{2}-\fr{\beta_ms_m}{2\sqrt{3}})X_\mu \right. ,\crn 
&\left. -\fr{gs_m}{\sqrt{3}}W^8_\mu + \fr{g_X}{\sqrt{6}}(-\fr{1}{2}-\fr{\beta_ms_m}{2\sqrt{3}})X_\mu
\right), \quad s_m = (-1)^m, \quad m=1,2, 
\end{align}
and we have used $X_m = -1/2 - \beta_ms_m/(2\sqrt{3})$ for left-handed lepton multiplets. 

As in \bib{Buras:2014yna}, we define the couplings between the $Z_1$ and $Z_2$ bosons to 
the fermions as follows
\begin{align}
\Delta_{m,k}^{ff}(Z_1) &= c_{m,\xi} \Delta_{m,k}^{ff}(Z) + s_{m,\xi} \Delta_{m,k}^{ff}(Z'),\crn
\Delta_{m,k}^{ff}(Z_2) &= -s_{m,\xi}\Delta_{m,k}^{ff}(Z) + c_{m,\xi} \Delta_{m,k}^{ff}(Z'),
\label{eq:Z_couplings}
\end{align}
where $k=L,R,A,V$ with $\Delta_V = \Delta_L + \Delta_R$, $\Delta_A = \Delta_R - \Delta_L$, with the following 
convention
\begin{align}
\mathcal{L}_f(Z,Z') &= \bar{f}\gamma^\mu [\Delta^{ff}_L(Z) P_L + \Delta^{ff}_R(Z) P_R] f Z_\mu\crn
& + \bar{f}\gamma^\mu [\Delta^{ff}_L(Z') P_L + \Delta^{ff}_R(Z') P_R] f Z'_\mu.
\end{align}
%%%
\begin{table}[h]
\begin{tabular}{|c|c|c|c|c|c|}
\hline
Model & Lepton & $\Delta_{L}(Z)$ & $\Delta_{R}(Z)$ & $\Delta_{L}(Z')$ & $\Delta_{R}(Z')$ \\
\hline
$M_1$     & $\nu$  & $\fr{g}{2c_W}$  & $0$             & $\fr{h_1g}{2c_W}\fr{1-(1+\sqrt{3}\beta_1)s_W^2}{\sqrt{3}\sqrt{1-(1+\beta_1^2)s_W^2}}$  & $0$\\
%\hline
$M_2$     & $\nu$  & $\fr{g}{2c_W}$  & $0$             & $-\fr{h_2g}{2c_W}\fr{1-(1-\sqrt{3}\beta_2)s_W^2}{\sqrt{3}\sqrt{1-(1+\beta_2^2)s_W^2}}$ & $0$\\
\hline
%\hline
$M_1$     & $e$    & $\fr{g}{2c_W}(2s_W^2-1)$  & $\fr{g}{c_W}s_W^2$  & $\fr{h_1g}{2c_W}\fr{1-(1+\sqrt{3}\beta_1)s_W^2}{\sqrt{3}\sqrt{1-(1+\beta_1^2)s_W^2}}$  & $\fr{h_1g}{c_W}\fr{-\beta_1 s_W^2}{\sqrt{1-(1+\beta_1^2)s_W^2}}$ \\
%\hline
$M_2$     & $e$    & $\fr{g}{2c_W}(2s_W^2-1)$  & $\fr{g}{c_W}s_W^2$  & $-\fr{h_2g}{2c_W}\fr{1-(1-\sqrt{3}\beta_2)s_W^2}{\sqrt{3}\sqrt{1-(1+\beta_2^2)s_W^2}}$ & $-\fr{h_2g}{c_W}\fr{\beta_2 s_W^2}{\sqrt{1-(1+\beta_2^2)s_W^2}}$\\
\hline
\end{tabular}
\caption{Couplings between the $Z$, $Z'$ gauge bosons and the left-, right-handed Standard Model's leptons in the models $M_1$ and $M_2$.}
\label{tab:Z_Z'_leptons_couplings}
\end{table}
Results for these couplings are given in \tab{tab:Z_Z'_leptons_couplings} for the case of the SM leptons. 
With the convention $h_2 = +1$ as in \refs{CarcamoHernandez:2005ka,Martinez:2014lta}, 
those $llZ$ and $llZ'$ couplings agree with \refs{CarcamoHernandez:2005ka,Martinez:2014lta}. 

In comparison with \bib{Buras:2014yna} we have to choose $h_1 = h_2 = +1$ to get the same sign 
for $s_{m,\xi}$. We agree with them for model $M_1$. 
For model $M_2$, which they call $F_2$, we can only agree if a minus sign is added to the 
$llZ'$ couplings.
 
It is important to note that the physical results 
such as the $e^+ e^- \to \mu^+ \mu^-$ cross section  
are independent of $h_m$ because it 
occurs both in the 
$\Delta_k^{ll}(Z')$ couplings and in $s_{m,\xi}$. The $Z-Z'$ interference terms 
are independent of $h_m$. Using the convention $h_1 = -h_2$, the above results show 
that the $llZ$, $llZ'$ couplings, $s_{m,\xi}$, and $c_{m,\xi}$ are the same in both models.

\bibliographystyle{h-physrev}
\bibliography{main}
\end{document}